\documentclass[12pt]{article}
\usepackage{latexsym,amsmath,amssymb,amsbsy,graphicx}
\usepackage[cp1251]{inputenc}
\usepackage[T2A]{fontenc}
\usepackage[space]{cite} 
\usepackage[centerlast,small]{caption2}%

\makeatletter\def\@biblabel#1{\hfill#1.}\makeatother

\allowdisplaybreaks \multlinegap0pt 

\begin {document}

{\flushleft {\itshape  Astrophysics }(the English translation of {\it Astrofizika}){\it, Vol. 59, pp. 475---483,
2016}\\
{{\texttt{DOI:}} 10.1007/s10511-016-9450-9}\\}
\bigskip
{\large{\bfseries ORIGIN OF THE BLUE CONTINUUM RADIATION IN THE
{\par \hspace*{3.5cm} FLARE SPECTRA OF dMe STARS\footnote{{The brief version of this article has been published in the Proceedings of the conference in honor of the 100th birthday
of Academician V. V. Sobolev (St. Petersburg, September 21}---25, 2015), pp. 219---221.}}}}

\bigskip
{\par \hspace*{0.5cm}\large{\bfseries E. S. Morchenko}$^{a,b}$}
\bigskip
{\par \hspace*{0.1cm}{ {\it Calculations of the emission
spectrum of a homogeneous plane layer of pure hydrogen plasma taking
into account nonlinear effects (the influence of bremsstrahlung and
recombination radiation of the layer itself on its Menzel factors)
\cite{Mor15}
show that the blue component of the optical
continuum  during the impulsive phase of large flares on dMe stars
originates from the near-photospheric layers \cite{Grin77}. The gas behind the front
of a stationary radiative shock wave propagating in the red dwarf chromosphere toward the photosphere is not capable of generating
the blackbody radiation observed at the maximum brightness of the flares.
 }
\par \hspace*{0.5cm} Keywords: {\it red dwarf stars: flares:
models of flares: plane layer: impulsive heating}}

\bigskip

 { \bfseries 1. Introduction.} Grinin and Sobolev \cite{Grin77}
 were the first to show that the optical conti-\\
nuum
 emission during the impulsive phase of large flares on dMe stars
 arises in  the ``transition region between the chromosphere and the
 photosphere.'' The near-photospheric layers  are heated by
 beams of {\it high-energy} ($\approx10\,\,\text{MeV}$) protons
 \cite{Grin_88,Grin_89} or/and ($\approx100\,\,\text{keV}$) electrons
 \cite{Grin_93}. The initial energy fluxes (``at the upper boundary of the flare
 region'' \cite{Grin_88}) in the proton and electron beams are
 $F_0\approx10^{11}-10^{12}\,\,\text{erg}\,\text{cm}^{-2}\,\text{s}^{-1}$ and
$F_0\approx3\cdot10^{11}\,\,\text{erg}\,\text{cm}^{-2}\,\text{s}^{-1}$,
respectively.

Katsova {\it et al.} \cite{Katsova_81} and Livshits {\it et al.} \cite{Livshits_81}\footnote{In the part examining the nature of the optical continuum of
stellar flares, the study \cite{Livshits_81} gives a brief discussion of work of {\underline{Katsova}} {\it et al.} \cite{Katsova_81}.} were the first to examine the
hydrodynamic response of the chromosphere of  a {\it red dwarf} to
impulsive heating by a high-power beam of accelerated electrons (a
low-energy cutoff $E_1=10\,\,\text{keV}$, a spectral index
$\gamma=3$,
$F_0=10^{12}\,\,\text{erg}\,\text{cm}^{-2}\,\text{s}^{-1}$). In solving
a single-temperature\footnote{$T_e=T_{i}$, where
$T_{e}$ is the electron temperature and $T_i$ is the ion one.}
system of gas dynamics equations ``with the given
boundary and initial conditions and the calculated loss and heating
functions`` \cite{Katsova_81}, they  found that two disturbances ``propagate
downward and upward`` \cite{Katsova_81} ``from the formed zone of
 high pressure`` \cite{Livshits_81}. The disturbance propagating toward the
photosphere (``downward'' \cite{Katsova_81}) ``is described in
subsequent times by a solution of the type of the second-kind
temperature wave \cite{Vol_63}.`` ``The latter is characterized by
a subsonic-velocity propagation of the thermal wave [temperature {\it jump}]`` \cite{Livshits_81}, ``in front of which a [{\it non-stationary}] shock wave ...`` develops \cite{Katsova_81}.

In these papers \cite{Katsova_81,Livshits_81} it was assumed that a region of
thickness $\Delta{}z\approx10\,\,\text{km}$ between the temperature
jump and the shock front (referred to below as a {\it chromospheric
condensation}) is the source of quasi-blackbody radiation at wavelengths
around 4500 \AA{} (see Fig. 4 in \cite{Katsova_81} and identical Fig. 8 in \cite{Livshits_81}). Katsova {\it et al.} \cite{Katsova_81}
and Livshits {\it et al.} \cite{Livshits_81}
note (pp. 162---163 in \cite{Katsova_81} and p. 281 in \cite{Livshits_81}) that the physical parameters of the
chromospheric condensation
($N_{\mathrm{H}}\approx2\cdot10^{15}\text{ cm}^{-3}$,
$T_{e}=T_{ai}=T\approx9000\text{ K}$, where $T_{ai}$ is the {\it atom-ion} temperature) are in the range of the parameters of a
{\it plane layer} in the model
of Grinin and Sobolev \cite{Grin77}
($N_{\mathrm{H}}\sim10^{15}-10^{17}\text{ cm}^{-3}$,
$T\sim{5000-20000}\text{ K}$,
$\Delta{z}\sim10^6-10^8\text{ cm}$). Here $N_{\mathrm{H}}$ is
equal to the sum of the hydrogen atom and proton concentrations.
 The chromospheric condensation \cite{Katsova_81},
however, lies at a height of $\approx1500\text{ km}$ \cite{Katsova_81,Livshits_81} above the level of the
quiescent photosphere of a red dwarf, {\it  i.e.,} much
higher than the homogeneous plane layer \cite{Grin77}.

The non-stationary  shock wave \cite{Katsova_81} propagates in the
{\it partially ionized} gas of the chromosphere of a red dwarf toward the
photosphere at a velocity of up to $\sim$ 100 km/s (see  Eqs. (A1), (A2) and Fig. 7 in \cite{Livshits_81}). The electron and {\it atom-ion} components of the chromospheric
plasma are heated {\it differently} behind the shock
front \cite{Katsova_81},\footnote{The velocities of the flow  are {\it subsonic} for
electrons, but supersonic for the ions and atoms
\cite{Pik_54}.} {\it i.e.,} it is true that the atom-ion temperature of
the post-shock gas \cite{Katsova_81} is considerably higher than its electron
temperature:
\begin{equation}
{\bf T_{ai}\gg{}T_e}
\end{equation}
(with the exception of the {\it dense} layers near the photosphere \cite{Grin77}). As a
result, the region between the temperature jump and the shock front \cite{Katsova_81}
is, in fact,  a {\it two-temperature region}\,.

In \cite{Mor15} the emission spectrum of a  {\it two-temperature}
($6\text{ eV}\leq{T_{ai}}\leq12\text{ eV}$,
$0.8\text{ eV}\leq{T_{e}}\leq1.5\text{ eV}$) {\it motionless} homogeneous {\it plane layer}
of pure hydrogen plasma ($3\cdot10^{14}\text{ cm}^{-3}\leq{}N_{\mathrm{H}}\leq3
\cdot10^{16}\text{ cm}^{-3}$) has been calculated
taking into account the influence of bremsstrahlung and recombination radiation of
the layer itself on its Menzel factors. This layer is located behind the
front of a {\it stationary plane-parallel} radiative shock wave (we use a frame of reference in which the shock front is at rest). In \cite{Mor15} the value of
$N_{\mathrm{H}}$ at which the intensity of the continuum spectrum
approaches the Planck function was determined.

From the start we have assumed that the optical depth in the
resonance transition in the center of the layer, $\tau_{12}^D$, is
approximately equal to $10^7$. The corresponding
layer thickness, $\mathcal{L}$, was introduced according to this condition (p. 2 and Eq. (53) in \cite{Mor15}).   The
transition from a transparent gas to a gas whose continuum emission
is close to the Planck function was, however, subsequently examined
for   the layer of a {\it fixed}  thickness $\mathcal{L}$ (see the first paragraph in section 7 in \cite{Mor15}). We have also assumed (the last paragraph in section 8) that
radiative cooling of the partially ionized gas heated at the front
of a {\it stationary} shock wave  can create a  zone
that is responsible for the blue component of the optical continuum
of  stellar flares.

The present paper shows that the blackbody radiation observed at the
maximum brightness of strong flares on dMe stars (the blue component of the optical continuum)
originates from  the deep (near-photospheric) layers
\cite{Grin77}. Various approaches for explaining the spectral
observations are analyzed for this purpose: the model of
Grinin and Sobolev \cite{Grin77}, the model of a second-kind
temperature wave \cite{Katsova_81,Livshits_81}, and the model of a
{\it plane-parallel} radiative shock wave propagating with a constant
velocity in the red dwarf chromosphere toward the photosphere.

{ \bfseries 2. Continuous spectrum.} We will show that the calculated
intensities of continuous radiation  from two- \cite{Mor15} and
single-temperature \cite{Grin77} homogeneous plane layers are similar
if:

(a) the electron temperature lies in the range $0.8
\text{ eV}\leq{}T_e\leq1.5\text{ eV}$, and

(b) the optical depth at a resonance transition is
\begin{equation}\label{Eq2}
\tau_{12}^D=\cfrac{k_{12}^D\mathcal{L}}{2}\,\sim10^7
\end{equation}
or more. Here $k_{12}^D$ is the absorption coefficient at the center
of the Doppler core of the $L_{\alpha}$ line (the value of
$k_{12}^D$ is defined by Eq. (53) in \cite{Mor15}).

Since $\tau_{12}^D\gg1$, photons leave the plasma in the far wings
of the spectral line (p. 6 in \cite{Mor15}). For this reason the mean escape
probability, $\theta_{12}$, for a photon of the resonance transition is
independent of the {\it atom-ion} temperature of the gas:
\begin{equation}
\theta_{12}\approx\left(\cfrac{\mathcal{B}_{21}\mathcal{E}_0}{\Delta{}\omega_{21}^D}\right)^{3/5}\cfrac{1}{(\tau_{12}^D)^{3/5}}\,,\,\,\,\,
\tau_{12}^D\propto\cfrac{1}{\sqrt{\pi}\Delta{}\omega_{21}^D}\propto{}T_{ai}^{-1/2}
\end{equation}
(see Eqs. (55), (53), and (43) in  \cite{Mor15}). Here $\mathcal{B}_{21}$ is the
corresponding Stark broadening parameter, $\mathcal{E}_0$ is the
Holtsmark field strength, and $\Delta{}\omega_{21}^D$ is the Doppler
half-width. Thus, the Menzel factors   and, therefore, the intensity of the continuous radiation from the layer \cite{Mor15} do not explicitly depend on the atom-ion
temperature.\footnote{A {\it weak} 
dependence of the Menzel factors on
$T_{ai}$ is caused by the parameter $b_{kn'}$ in Eq. (56) of study
\cite{Mor15}.}

Solution of the balance equations for the populations of the levels
shows \cite{Mor15} that the Menzel factors  differ little from
unity so that the source function $S_\nu\approx{}B_{\nu}(T_e)$. This
result remains true when $T_{ai}=T_e=T$. Grinin and Sobolev
\cite{Grin77} begin with the assumption of thermodynamic equilibrium
in the dense gas of a stellar flare: $S_{\nu}=B_\nu(T)$ (Eq. (9)
in \cite{Grin77}).

Taking into account that the Gaunt factors for bremsstrahlung and
photoionization absorption in the optical range are on the order of
unity ({\it e.g.,} \cite{iva69}, pp. 16---18) and that the number $k_{\mathrm{max}}$ of attainable atomic
levels is high,\footnote{For an electron density
$N_e\approx10^{15}\text{ cm}^{-3}$, the number  of the level
defining the boundary of the visible continuum  is 13 (the Inglis-Teller relation).} we find that
the radiation intensities calculated using Eq. (25) in \cite{Mor15} and Eq.
(9) in \cite{Grin77} are  close, with
\begin{equation}
I_{\nu}\approx{}B_{\nu}(T)\cdot(1-e^{-\tau_\nu}),
\end{equation}
where
\begin{equation}
\tau_{\nu}=\varkappa_{\nu}^{(b)}\mathcal{L}\left[1+2\cfrac{\mathrm{Ry}}{T}\sum\limits_{j=k}^{k_{\mathrm{max}}}\cfrac{e^{\beta_j}}{j^3}\right],\,\,
\beta_j\equiv\cfrac{\mathrm{Ry}}{j^2T}\,,
\end{equation}
$\varkappa_{\nu}^{(b)}$ is the bremsstrahlung absorption coefficient
(see Eq. (29a) in \cite{Mor15}). The sum is taken
over all levels for which the threshold photoionization frequency
$\nu_j\leq\beta_kT/h$; then  $h\nu\geq\beta_kT$.\footnote{It is
evident that the exponent ``39'' in Eq. (4) of study \cite{Grin77} should read
``-39;'' and the temperature $T$ should be raised to a power of
``-1/2.''}

Katsova {\it et al.} \cite{Katsova_81} and Livshits {\it et al.} \cite{Livshits_81} give the following physical parameters of the source of quasi-blackbody radiation at wavelengths
around 4500 \AA{}: $\Delta{z}\approx10\text{ km}$, $N_{\mathrm{H}}\approx2\cdot10^{15}\text{ cm}^{-3}$,
$T\approx9000\text{ K}\sim
0.8\text{ eV}$. It is easy to see that the
corresponding {\it single-temperature} ($T_{ai}=T_e$)  homogeneous plane layer cannot generate the small Balmer jump
(the bottom curve of Fig. 1).\footnote{Here, as in \cite{Katsova_81,Livshits_81}, we use the model of a homogeneous plane layer to calculate
the radiation intensity from the chromospheric condensation \cite{Katsova_81}. The validity of such approach is not discussed in this paper.} Evidently, this layer  also cannot
explain the blue continuum radiation observed at the maximum brightness of strong flares, since the gas inside the layer \cite{Katsova_81} is {\it transparent}
in the continuous spectrum  (the optical depth at
wavelength  $\lambda=4170\,\text{\AA}$ is
$\tau_{4170}\sim0.02\ll1$ \cite{Mor15}).
\begin{centering}
\begin{figure}
\includegraphics[width=11cm]{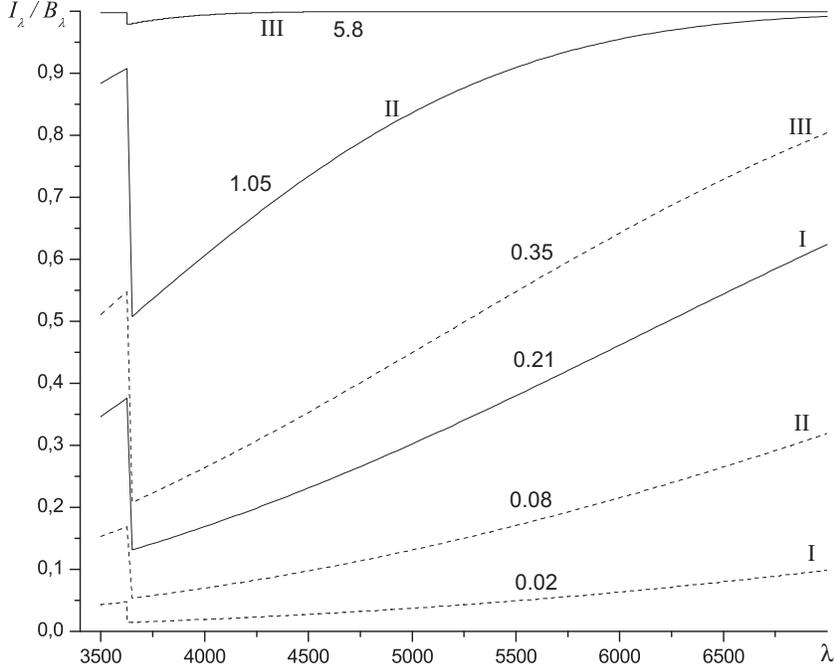}
\caption{{{The emission spectrum of a homogeneous plane layer of
thickness $\mathcal{L}=10\text{ km}$ \cite{Mor15} for
$T_{ai}=T_e=T$. The dashed curves correspond to $0.8\text{ eV}$ and
the smooth curves, to $1\text{ eV}$. The Roman numerals correspond
in increasing order to $N_{\mathrm{H}}=2{\cdot}10^{15}
\text{ cm}^{-3}$, $N_{\mathrm{H}}=7{\cdot}10^{15}\text{ cm}^{-3}$, and
$N_{\mathrm{H}}=3{\cdot}10^{16}\text{ cm}^{-3}$.}}}
\end{figure}
\end{centering}
 On the other hand, emission from a
{\it denser} gas with the same layer thickness can explain well  both the
observed color indices of strong flares
and the small size of the Balmer jump  \cite{Grin77}.

In order to {\it verify} these statements, let us calculate the value of $\tau_{4500}$  for a homogeneous plane layer
with $S_{\lambda}=B_{\lambda}(T)$.
 Table 1 shows the results. The equilibrium electron concentration, $N_e^{\mathrm{eq}}$, is calculated
according to  the Saha equation. The values of $k_{\mathrm{max}}$ and $\tau_{4500}$ are obtained from Eq. (12) in \cite{Mor15}  and Eq. (5) of the present article, respectively.
From this table, it follows that the layer with $N_{\mathrm{H}}=2{\cdot}10^{15}\text{ cm}^{-3}$, $T=9000\text{ K}$
generates the blackbody-like
continuum in the blue-visible region of the spectrum  at $\mathcal{L}\gtrsim215\,\,\text{km}$.\footnote{
$\tau_{4500}\approx4.7\cdot10^{-3}\cdot215\approx1$.}

\begin{table} \small \caption{Physical parameters of the plasma inside the layer with $S_\lambda=B_{\lambda}(T)$}
\begin{tabular}{cccccc}
\hline $N_{\mathrm{H}},\,\,\text{cm}^{-3}$ &$T,\,\,\text{K}$ &${\mathcal L},\,\,\text{km}$ & $N_e^{\mathrm{eq}},\,\,\text{cm}^{-3}$ & $k_{\mathrm{max}}$ &$\tau_{4500}$\\
$10^{15}$ & 9000 &10 &$1.99\cdot{10^{14}}$ &16&$2.2\cdot{10^{-2}}$ \\
$2\cdot{}10^{15}$ & 9000&10 &$2.91\cdot10^{14}$ &15 &$4.7\cdot10^{-2}$\\
$10^{16}$ &9000&10 &$6.8\cdot{10^{14}}$ &14 &0.255\\
$10^{17}$ &9000 &10 &$2.2\cdot{10^{15}}$ &12&2.65\\
\hline
\end{tabular}
\end{table}

In the model \cite{Katsova_81} the formation of the chromospheric
condensation with  the necessary thickness $\Delta{}z\approx10\text{ km}$ occurs only when the
shock front enough widely separates from the temperature jump (see  Fig. 8 in \cite{Livshits_81}).  At earlier times the geometric thickness of the gas layer
 \cite{Katsova_81}  $\Delta{z}<10\,\,\text{km}$, since the temperature jump moves at a {\it subsonic} velocity. The plane layer with a smaller
thickness $\mathcal{L}$ is still more transparent in the optical continuum.

We note that for the layer with  $T=0.8\text{ eV}$ and $\mathcal{L}=10\text{ km}$, a comparatively
small Balmer jump can be obtained only for a {\it near-photospheric}
concentration $N_{\mathrm{H}}=3\cdot10^{16}\text{ cm}^{-3}$ (see the
lower curve labeled III in Fig. I). A slight increase in temperature
($T=1\text{ eV}$) leads to a reduction in the required concentration (see
the top curve for $N_{\mathrm{H}}=7\cdot10^{15}\text{ cm}^{-3}$).

The  homogeneous plane layer with
$N_{\mathrm{H}}=3\cdot10^{16}\text{ cm}^{-3}$, $T=1\text{ eV}$,
and $\mathcal{L}=10\text{ km}$ provides the blue continuum
radiation from a stellar flare (the top curve in Fig. 1). Fig. 2 also demonstrates
that for $T=1.2\text{ eV}$,
$N_{\mathrm{H}}=3\cdot10^{16}\text{ cm}^{-3}$, and the layer
thickness $\mathcal{L}=20\text{ km}$, the emission lines are
entirely ``sunk'' in the continuous spectrum. These concentrations, temperatures, and thicknesses of the blackbody radiation
sources lie within the range of parameters for a pure hydrogen plasma
layer in the model of Grinin and Sobolev \cite{Grin77}.   In a study of the fine time structure of two flares
on AD Leo, Lovkaya \cite{lov13} has found that at the maximum
brightness these two flares radiated as absolute black bodies with
temperatures of approximately $1.2$ and $1.12\text{ eV}$. Based on a detailed
colorimetric analysis, Zhilyaev {\it et al.} \cite{zhil12} have determined
a temperature $T\approx1.16\,\text{eV}$ for the blackbody radiation
at the peak of a strong flare on the red dwarf EV Lac. The authors \cite{Grin77} have found (p. 354) that the homogeneous plane layer with
$N_{\mathrm{H}}=10^{16}\text{ cm}^{-3}$,  $\Delta{}z=1.5\cdot10^6\text{ cm}$, and $T=11000$\text{ K} generates the
 energy distribution in the continuum, which is in
close agreement with the energy distribution in the continuous spectrum at the maximum brightness of another flare on EV Lac.

\begin{figure}\label{Fig1}
\includegraphics[width=11cm]{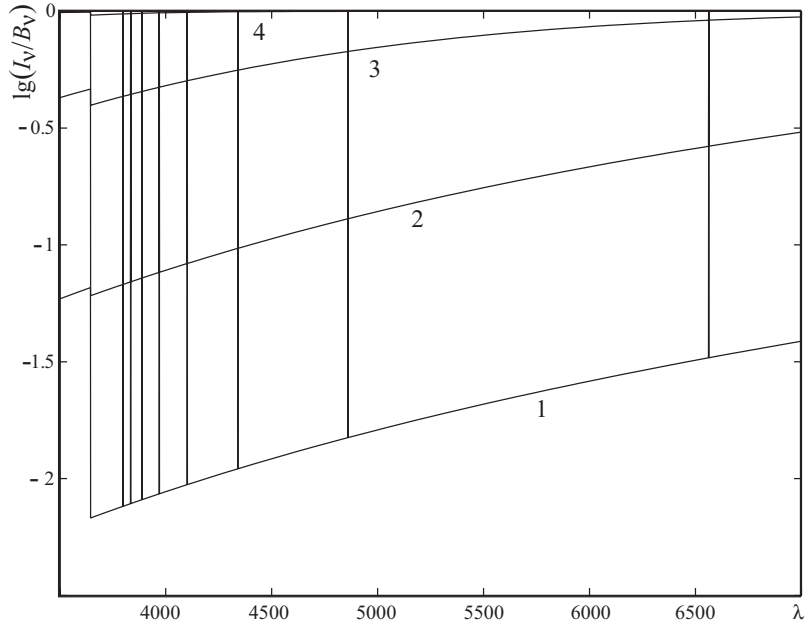}
\caption{{ The emission spectrum of a homogeneous plane layer of
thickness $\mathcal{L}=20\text{ km}$ \cite{Mor15} with
$T_{ai}=T_e=T=1.2\text{ eV}$. The uppermost curve (labeled 4)
corresponds to $N_{\mathrm{H}}=3\cdot10^{16}\text{ cm}^{-3}$.}}
\end{figure}

Therefore, a homogeneous plane layer with the parameters corresponding
to the chromospheric condensation  \cite{Katsova_81} (the model of {\underline{Katsova}} {\it et al.}), as opposed to a {\it dense} layer
in the model of Grinin and Sobolev \cite{Grin77}, {\it cannot explain} the continuous spectrum of
stellar flares.\footnote{See also {\it Appendix} in the present article.}

{\bfseries 3. Stationary radiative shock wave.} In \cite{Mor15} it was assumed that radiative cooling of the gas behind the front of
a {\it stationary} plane-parallel shock wave propagating in the red dwarf chromosphere toward the photosphere  is capable of creating a zone with near-photospheric
concentration $N_{\mathrm{H}}=3\cdot10^{16}\text{ cm}^{-3}$ (this value corresponds to the value of $N_{\mathrm{H}}$ of the source of the blue continuum radiation at
$T_e=1\text{ eV}$
in the framework of the homogeneous plane layer model). Here we
clarify the conditions under which this is possible.

Fadeyev and Gillet \cite{fad01}, Bychkov {\it et al.} \cite{BBMN2014}
have calculated the profile of a stationary plane-parallel {\it radiative} shock wave with
detailed accounting for elementary processes in the post-shock
plasma: the electron impact ionization, the triple recombination, the electron
impact excitation and de-excitation, etc.\footnote{{In \cite{BBMN2014} the numerical calculations
have been performed for  two-levels atoms and ions.}}  The following parameters
were chosen in \cite{BBMN2014} for the unperturbed gas: the total concentration of
ions and atoms $N_0=10^{12}\,\text{ cm}^{-3}$, a temperature
$T_0=3000\text{ K}$, and a {\it magnetic field} $H_0=2\text{ G}$.  The magnetic field is
oriented perpendicular to the gas velocity \cite{BBMN2014}. The
plasma flows through the discontinuity surface at a velocity of
$u_0=60\text{ km/s}$ (we use a frame of reference in which the shock front is at rest).

Let us take $N_0=10^{14}\text{ cm}^{-3}$  for the unperturbed
chromosphere of a dMe star and retain the other parameters of the
gas {\it without modification}. Immediately after the discontinuity for
the same shock velocity $u_0$ we have:
$N_1\approx3.9\cdot10^{14}\text{ cm}^{-3}$,
$T_{ai1}\approx1.03\cdot10^5\text{ K}\approx8.86\text{ eV}$,
$T_{e1}\approx7.42\cdot10^3\text{ K}\approx0.64\text{ eV}$, and
$H_1\approx7.8\text{ G}$ (the atoms and ions are
heated at the shock front along the Rankine-Hugoniot adiabat and
the electrons, along the Poisson adiabat). Here the ionization state of the unperturbed gas is calculated
via the solution of the system of the Saha equations; the ionization
of {\it metals} is taken into account  (for more details, see pp. 651---653 in \cite{BBMN2014}). Clearly,
${\bf T_{ai1}\gg{}T_{e1}}$ (see also paragraph 4 in the introduction to
this article). In these calculations we use a reduced value of the pre-shock gas
temperature ($3000\text{ K}$ instead $\sim5000-8000\text{ K}$). In the
present problem, however, the Mach number is quite high
($M_0={u_0}/{v_{s0}}\approx10.3\gg1$, where $v_{s0}$ is the adiabatic
sound speed), so the time ``history'' of radiative cooling of the post-shock gas should not depend too strongly
on the background temperature $T_0$.

Calculations \cite{BBMN2014} show that the optical depth in the
$\mathrm{Ly-\alpha}$ line in the {\it cooling region} of the post-shock gas is
$\tau_{12}\sim10^7$ ($\tau_{12}$ is reckoned from the discontinuity surface).
The authors \cite{Mor15} have shown that $\tau_{12}$ depends
weakly on the  concentration of the hydrogen atoms in the post-shock plasma (see Eqs. (2)---(3) in \cite{Mor15}).
Thus, under the conditions of the chromosphere of a red dwarf, we
can set $\tau_{12}\sim10^7$ in a first approximation (as in section 2 of study \cite{Mor15}).

A {\it two-temperature} ($T_{ai}>{}T_e$) {\it plane} layer of thickness
$\mathcal{L}$ can be treated as the \emph{simplest} approximation
for the radiative cooling region behind the front of a stationary {\it plane-parallel} shock wave. For
concreteness we assume that $\tau_{12}$ is the optical depth at the
point $\mathcal{L}/2+l$, where $l$ is the  geometric
size of the region where the electron temperature $T_e$ {\it is raised} by
the elastic collisions of electrons with ions and atoms \cite{fad01,BBMN2014}. Since
$\mathcal{L}\gg{}l$ \cite{fad01,BBMN2014}, we will assume that
$\tau_{12}\approx\tau_{12}^D$ (see also the paragraph six of
introduction to the present article).

With increasing $N_{\mathrm{H}}$ in the cooling region of the post-shock plasma, the {\it geometric thickness}
$\mathcal{L}$ of the emitting plane layer {\it decreases}. In fact, for
$N_{\mathrm{H}}=3\cdot10^{16}\text{ cm}^{-3}$, $T_e=1\text{ eV}$,
and $T_{ai}=8\text{ eV}$ (an estimate), according to Eq.
(\ref{Eq2}) we obtain that
\begin{equation}\label{Eq6}
\mathcal{L}=\cfrac{2\tau_{12}^D}{k_{12}^{D}}\sim2\cdot10^7\cdot\left[4\pi^{3/2}\sqrt{\cfrac{m_{\mathrm{H}}}{m_e}}\,a_0^2\left(\cfrac{\mathrm{Ry}}{E_{21}}\right)f_{12}\sqrt{\cfrac{\mathrm{Ry}}{T_{ai}}}%
\,N_{\mathrm{gr}}\right]^{-1}\approx500\,\text{m}.\footnote{It is assumed here that there is no connection between the increasing in $N_{\mathrm{H}}$ and
the change in the magnetic field $H$ during all non-stationary radiative cooling of the post-shock gas.}
\end{equation}
Here $m_{\mathrm{H}}$ is the hydrogen atom mass, $m_e$ is the electron mass, $a_0$ is the Bohr radius, $E_{21}$ is the excitation energy of the second level of the
hydrogen atom, $f_{12}\approx0.416$ (the absorption oscillator
strength for the transition $1\rightarrow2$), and $N_{\mathrm{gr}}$ is the concentration of atoms in the
ground state (here, for concreteness, we assume that $N_{\mathrm{gr}}\approx2\cdot10^{16}\text{ cm}^{-3}$). It is clear that $\mathcal{L}\propto\sqrt{T_{ai}}$, so
that a reduction in the atom-ion temperature in Eq. \eqref{Eq6} leads to a
{\it drop} in $\mathcal{L}$.

According to study of Lovkaya \cite{lov13}, the linear sizes of the flares on AD Leo at the maximum brightness are approximately
${ 10^9}\text{ cm}$ (see p. 609 in \cite{lov13}). Since $\mathcal{L}$ is \emph{much less} than  $10^9\text{ cm}$, our
assumption \cite{Mor15} regarding the origin of the blue continuum radiation
 {\it is not confirmed}.

{\bfseries 4. Additional remarks.} Sobolev and Grinin \cite{sob95}\footnote{ The brief version of this article is available
via the ADS Article Service in the Proceedings of Cool Stars 9: {\texttt {http://adsabs.harvard.edu/abs/1996ASPC..109..629S.}}} assume
that the line  spectrum of stellar flares is formed in the chromospheric layers where
``gas-dynamic effects caused by the rapid release of energy play an
extremely important role \cite{Katsova_81}.''  In the present article
it has been shown that the homogeneous plane layer of pure hydrogen
plasma with the parameters obtained by Katsova {\it et al.} \cite{Katsova_81}
is transparent in the optical continuum.  This result is in agreement with \cite{sob95}. Thus, the model \cite{Katsova_81} can explain ({\it qualitatively}) the
increased intensity of the hydrogen emission
lines in the spectra of stellar flares.

In the gas-dynamic calculations  \cite{kow15,kow15arhiv} Kowalski
 has increased the energy flux in the
electron beam to
$F_0=10^{13}\text{ erg}\,\text{cm}^{-2}\,\text{s}^{-1}$  (a low-energy cutoff
$E_1=37\text{ keV}$). The author \cite{kow15arhiv}
believes that ``the ...  optical continuum radiation [during the impulsive phase of stellar flares]  originates from the
chromospheric condensation with [electron] densities as high as
$N_{e,\mathrm{max}}\approx5.6
\cdot10^{15}\text{ cm}^{-3}$ [,
$T\approx12800\text{ К}$] and from non-moving ... dense ($N_e\approx10^{15}\text{ cm}^{-3}$ [, $T\sim10.000\text{ K}$]) layers
below the chromospheric condensation'' \cite{kow15arhiv} (only $\approx25\%$ of the
blue continuum radiation is produced in these layers
\cite{kow15arhiv}).

The calculations of Grinin {\it et al.} \cite{Grin_93},\footnote{The full text of the article is available via the ADS Article Service:\\
 \texttt{http://adsabs.harvard.edu/abs/1993ARep...37..182G},
\texttt{http://adsabs.harvard.edu/abs/1993ARep...37..187B}.}
 which contradict the viewpoint \cite{kow15,kow15arhiv}, {\it were not discussed} in \cite{kow15,kow15arhiv}. As
justification for raising $F_0$ to
$10^{13}\,\,\text{erg}\,\text{cm}^{-2}\,\text{s}^{-1}$, it was pointed
out \cite{kow15} that ``non-thermal ... {\it deka-eV}\, electrons'' rapidly lose their energy
during interactions with the chromosphere of a red dwarf and, for
this reason, cannot  heat the deep layers of the star's
atmosphere. But in \cite{Grin_93}
the energy losses of {\it high-energy} electrons through ionization of atoms and Coloumb interactions
with free electrons are taken into account (see Eqs. (1) and (2) in \cite{Grin_93}).  In addition, $F_0=10^{13}\,\,\text{erg}\,\text{cm}^{-2}\,\text{s}^{-1}$ is two orders of
magnitude {\it greater} than the value of $F_0$ usually used in
gas-dynamic models of {\it solar} flares (see Somov {\it et al.} \cite{som81}).\footnote{Here we have in mind the
problem of  {\it return current}.}   At the same time,
with
$F_0=10^{11}-10^{12}\,\,\text{erg}\,\text{cm}^{-2}\,\text{s}^{-1}$
and the same beam parameters \cite{kow15arhiv}, it has not been possible to reproduce
the optical continuum spectrum of stellar flares \cite{kow15arhiv}.

Therefore, Kowalski's conclusion regarding the nature of $\approx75\%$ \cite{kow15arhiv} of  the blue
continuum radiation {\it is not sufficiently substantiated}.

The plasma behind the front of a stationary radiative shock wave moves at a
subsonic velocity relative to the discontinuity surface. In addition, as the
post-shock plasma radiates, the velocity of the gas flowing out of
the discontinuity, $u(t)$, {\it decreases} because the gas becomes denser. From the
standpoint of a laboratory observer, however, the velocity of the
gas, $u^l(t)$, is equal to $u_0-u(t)$. As a result, the line {\it core} of the emission line of the post-shock plasma should be shifted as a
whole in the direction of motion of the shock front. It is interesting to note
that the ${\mathrm{H}}_{\alpha}$ line profile in the spectrum of a flare on
dM5.6e (see Fig. 7a in Eason {\it et al.} \cite{eas}) has a Doppler core that is
shifted to the blue, rather than red (see also subsection 5.3 in \cite{Mor15}). The gas behind the front of a shock wave propagating in the {\it partially
ionized } chromosphere
{\it upward} can generate this radiation.

{\bfseries 5. Model of a second-kind temperature wave.} Katsova {\it et
al.} \cite{Katsova_81} and Livshits {\it et al.} \cite{Livshits_81} do not take into account the inequality
$\bf{}T_{ai}\gg{}T_e$  in the post-shock chromospheric plasma of stellar and {\it \underline{solar}} flares, and this is a {\it fundamental deficiency} of their models
\cite{Katsova_81,Livshits_81}. {\it{\bf As it follows from pp. 3666---3667 and  Fig. 1 in \cite{Katsova_15},}} {\underline{Katsova}} and Livshits {\it do not dispute}
this statement. (``We apply the two-temperature approximation where the electrons
and ions are heated differently behind the shock front'' (p. 3667 in \cite{Katsova_15}).\footnote{In order to
avoid misunderstandings, we recall that the ``opposite'' situation
occurs in {\it the EUV part} of solar and stellar flares; {\it i.e.,} the
electron temperature of the plasma  is considerable higher than
the {\it ion} temperature:  ${\bf T_e\gg{}T_i}$
 (see Somov {\it et al.} \cite{som81}).})

{\bfseries 6. Conclusion.} Let us summarize the astrophysical
results obtained in this paper.

 (a) It has been demonstrated that the blue continuum radiation observed at the maximum brightness of large
 flares on dMe stars originates from the near-photospheric layers \cite{Grin77}.

 (b) It has been shown {\it for the first time} that the homogeneous plane layer
 with the parameters corresponding to the chromospheric condensation
 \cite{Katsova_81}, as opposed to the {\it denser} layer in the model
 \cite{Grin77}, {\it cannot explain} the  optical continuum spectrum of  stellar flares.

Our conclusions are based on an analysis of three {\it different}
approaches  to explaining the spectral observations using the model of a homogeneous  plane
gas layer. In \cite{Grin77} this layer is located in the
``transition region between the chromosphere and the photosphere.''
In the gas-dynamic calculations \cite{Katsova_81}, a {\it single-temperature}
($T_e=T_{ai}=T$) plane layer corresponds to the chromospheric condensation formed by a
second-kind temperature wave.  Finally, in the model of a stationary
plane-parallel shock wave propagating in the partially ionized chromosphere of a red
dwarf toward the photosphere, a {\it two-temperature} ($T_{ai}\neq{}T_e$) gas layer
is the simplest approximation for the radiative cooling region. We
note that this model {\it differs} from the models \cite{Katsova_81,Livshits_81}. In
fact, Katsova {\it et al.} \cite{Katsova_81} and Livshits {\it{et al.}} \cite{Livshits_81}  introduce a ``system of [partial differential]
equations of one-dimensional gravitational gas dynamics'' \cite{Katsova_81}  taking into account the {\it thermal conductivity} (Fourier's law), while
the authors \cite{fad01,BBMN2014} consider a system of ordinary differential equations
with detailed accounting for {\it non-stationary} radiative cooling
in the post-shock plasma.\footnote{The thermal conductivity and the gravitational acceleration {\it are not taken into account.}}

{\small{{\bf  Acknowledgments.} The author thanks the reviewers for
constructive critical comments. I am also grateful  to T. V. Morchenko for assistance.  This work was supported by the Scientific School
(project code 1675.2014.2 NSh).

The present article is dedicated to the
memory of my dear physics teacher, Vladimir Mikhailovich Petrusenko
(1948---2016).}\\}

 Translated from {\it{Astrofizika}}, Vol. 59,  pp. 535---545, October---December, 2016. Original article submitted October 9, 2015; resubmitted
March 26, 2016; accepted for publication August 24, 2016. {\it This is a  revised and extended version.}

{\small {\flushleft  $^{a}$ Sternberg  Astronomical Institute,
 Lomonosov Moscow State University, Russia;\\
$^{b}$ Faculty of Physics  Lomonosov Moscow State University,
Russia.}}\\
{E-mail: {\texttt{morchenko@physics.msu.ru}}}.

\begin{center}
{\bfseries  APPENDIX\footnote{This part of the article is published only in {\texttt{astro-ph.SR}}.}}
\end{center}

{$\bullet$} To calculate the radiation intensity from the chromospheric condensation \cite{Katsova_81}, Katsova {\it et al.} \cite{Katsova_81}
and Livshits {\it et al.} \cite{Livshits_81} use the {\it modified} Eq. (9) from paper by Grinin and Sobolev \cite{Grin77}:
\begin{equation}
I_{\mathrm{fl}}=B_{\lambda}(T)\cdot(1-e^{-\Delta{\tau}_{\lambda}}),
\end{equation}
where the value of $\Delta{\tau}_{\lambda}$ is equal to
\begin{equation}
\int{}a(T,p_e)p_ed\xi;
\end{equation}
$p_e$ is the electron pressure, ``$a(T,p_e)$
is taken in accordance with the tables of \cite{Bode}, $\xi$ is a Lagrangian variable: $d\xi=-N_{\mathrm{H}}\,dz$'' \cite{Katsova_81}. The integration is
carried out ``over the complete
high-density region {\it for different times}'' (p. 163 in \cite{Katsova_81}).\footnote{Eq. (7)  corresponds  to that on p. 281 in \cite{Livshits_81};
see also Eq. (4) in \cite{Livshits_81}.}

Bode's tables \cite{Bode} contain data for  {\it homogeneous} plasma
whereas the chromospheric condensation \cite{Katsova_81} is rather inhomogeneous.
Therefore, Eq. (8) is incorrect. Moreover, this Eq. cannot be used in the calculations of the radiation intensity from a  {\it  plane layer}.
The  solution of this problem is the replacement of $\Delta{\tau}_{\lambda}$ in Eq. (7) with $\tau_{\lambda}$ (as in Table 1).

{$\bullet$} The physical parameters of the condensation \cite{Katsova_81} are close to the parameters of the {\it reversing layer} of an A0 star (see  Allen \cite{Allen}, \S\,\,103).
Using the absorption coefficient for such plasma \cite{Bode},  let us calculate the optical depth at wavelength 5000 \AA{}, ${\tau}_{5000}^{\mathrm{Bode}}$,
at $\Delta{z}=10\text{ km}$:
\begin{equation}
{\tau}_{5000}^{\mathrm{Bode}}\approx10^{0.97}\cdot\mu{}\mathrm{Ar}\cdot{N_a}\Delta{z}\approx0.03\sim10^{-2}\ll1.
\end{equation}
Here $\mu$ is the molecular weight of the gas without the electrons and metals ($\mu\approx1.235$ \cite{Allen}), $\mathrm{Ar}$ is the unified atomic mass unit, $N_a$ is the total
number density of atoms ($N_a\approx1.585\cdot10^{15}\text{ cm}^{-3}$ \cite{Allen}). One can see that ${\tau}_{5000}^{\mathrm{Bode}}$ coincides in order of magnitude
with  $\tau_{4500}$ (Table 1) at $N_{\mathrm{H}}=2\cdot
10^{15}\text{ cm}^{-3}$ and $T=9000 \text{ K}$. Note that
the value of $N_{\mathrm{H}}$ inside the chromospheric condensation \cite{Katsova_81} is {\it strictly less} than ${\bf 10^{16}}\text{ cm}^{-3}$ for all values of $\xi$ (in particular for
$\xi=10^{21}\text{ cm}^{-2}$ corresponding to $\Delta{z}=10^6\text{ cm}$ at $N_{\mathrm{H}}=10^{15}\text{ cm}^{-3}$ \cite{Katsova_81}).

{$\bullet$} Katsova {\it et al.} \cite{Katsova_81} (p. 163) and Livshits {\it et al.} \cite{Livshits_81} (p. 281)  state   that the optical radiation from the
 {\it plane layer} with
$N_{\mathrm{H}}=2\cdot10^{15}\text{ cm}^{-3}$, $T\approx9000\text{ K}$, $\Delta{}z=10\text{ km}$  provides the color indices, which
are ``in a good agreement with the observed blue-white radiation'' \cite{Livshits_81} of a stellar flare ($U-B=-1,\,\,B-V=0.5$ \cite{Katsova_81,Livshits_81}).

 In the present article it has been  demonstrated that
this  layer is {\it transparent} in the optical continuum.  As it follows from \cite{Grin77},
the opacity of the gas beyond the Balmer jump is a necessary condition for explaining the color indices at the maximum brightness of large stellar flares.  Therefore,
the {\it real} values of $U-B$ and $B-V$  for the layer \cite{Katsova_81} {\it are not in the blue-visible region}. In addition,
the published color indices \cite{Katsova_81} do not correspond the color indices at the peaks of strong flares: $U-B\approx-1.0$ and
$B-V\approx0.2$ (p. 300 in \cite{Grin_90}\footnote{\texttt{http://adsabs.harvard.edu/abs/1990IAUS..137..299G.}}); $U-B=-0.91\pm0.02$
and $B-V=0.0\pm0.02$ (Zhilyaev {\it et al.} \cite{zhil12}).

{$\bullet$} Again, Katsova {\it et al.} \cite{Katsova_81} (p. 163) and Livshits {\it et al.} \cite{Livshits_81} (p. 281) state that the layer \cite{Katsova_81}
provides the value of the Balmer jump $D\approx0.8$ (``in accordance with Grinin and Sobolev \cite{Grin77}'' \cite{Livshits_81}). Here the
value of the Balmer jump ``is determined by the formula
\begin{equation}
D=\lg\cfrac{I_{\nu>\nu_2}}{I_{\nu<\nu_2}}\,,
\end{equation}
where $I_{\nu>\nu_2}$ and $I_{\nu<\nu_2}$ are the radiation intensities immediately after and immediately before the jump'' \cite{Grin77}. In reality,
as it follows from Fig. 3,
\begin{figure}\label{Fig1}
\includegraphics[height=19cm,width=13cm]{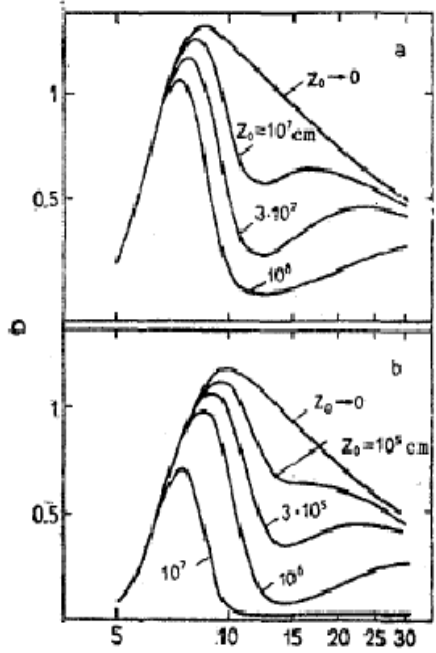}
\caption{``Balmer jump $D$ as function of the temperature T [($10^3\text{ К}$)] for different
thickness $z_0$ of the emitting region: a) $N_{\mathrm{H}}=10^{15}\text{ cm}^{-3}$, b) $N_{\mathrm{H}}=10^{16}\text{ cm}^{-3}$'' (Grinin and Sobolev \cite{Grin77}). In
 \cite{Grin77} the
emission of {\it negative hydrogen ions}  at $T<10^4\text{ K}$ is taken into account.}
\end{figure}
the value of $D\approx0.8$ for the layer with thickness $\sim10^8\text{ cm}\sim{\bf10^3}\text{ km}$
($T\approx9000\text{ K}$, $N_{\mathrm{H}}=
10^{15}\text{ cm}^{-3}$).
{\small

\newpage

}

\end{document}